\documentclass{mem}
\usepackage{amsmath}
\usepackage{natbib}
\usepackage{txfonts}
\usepackage{balance}
\usepackage{graphicx}
\usepackage[a4paper]{hyperref}
\idline{81}{00}

\usepackage{color}
\usepackage[normalem]{ulem}
\definecolor{lightred}{rgb}{0.8,0.0,0.0}
\definecolor{lighterred}{rgb}{0.8,0.6,0.6}
\definecolor{lightgreen}{rgb}{0.1,0.6,0.1}

\begin{document}

\title{
Numerical simulations of wave propagation in the solar chromosphere
}

\subtitle{}

\author{
C. \,Nutto
\and O. \, Steiner
\and M. \, Roth
          }

%\offprints{C. Nutto}
 
\institute{
Kiepenheuer-Institut f\"ur Sonnenphysik, Sch\"oneckstr. 6, 
79104 Freiburg, Germany\\ 
\email{nutto@kis.uni-freiburg.de}
}

\authorrunning{Nutto et al.}

\titlerunning{Numerical wave propagation in the solar atmosphere.}

\abstract{ We present two-dimensional simulations of wave propagation
  in a realistic, non-stationary model of the solar atmosphere. This
  model shows a granular velocity field and magnetic flux
  concentrations in the intergranular lanes similar to observed
  velocity and magnetic structures on the Sun and takes
  radiative transfer into account.

We present three cases of magneto-acoustic wave propagation through
the model atmosphere, where we focus on the interaction of different
magneto-acoustic wave modes at the layer of similar sound and Alfv\'en
speeds, which we call the equipartition layer. At this layer acoustic
and magnetic mode can exchange energy depending on the angle between
the wave vector and the magnetic field vector.

Our results show that above the equipartition layer and in all three
cases the fast magnetic mode is refracted back into the solar
atmosphere. Thus, the magnetic wave shows an evanescent behavior in
the chromosphere. The acoustic mode, which travels along the magnetic
field in the low plasma-$\beta$ regime, can be a direct consequence of
an acoustic source within or outside the low-$\beta$ regime, or
it can result from conversion of the magnetic mode, possibly from
several such conversions when the wave travels across a series of
equipartition layers.  \keywords{MHD -- waves -- Sun: atmosphere --
  Sun: chromosphere -- Sun: helioseismology -- Sun: photosphere} }

\maketitle{}

\section{Introduction}

Classical helioseismology relies on waves with frequencies below the
acoustic cut-off frequency. These type of waves are trapped inside the
acoustic cavity of the Sun. High frequency waves above the acoustic
cut-off frequency, however, are able to travel freely into the solar
atmosphere. Along their path through the atmosphere these waves
interact with the complex magnetic field that is present in the
photosphere and the chromosphere. Being able to observe these running
waves in the atmosphere with high cadence instruments like MOTH
\citep{nutto_finsterle04b} a new field in helioseismology is
emerging. However, the analysis tools at hand are still very basic
\citep{nutto_finsterle04, nutto_haberreiter07} and the
characterization of the interaction of the magneto-acoustic waves with
the magnetic field is still difficult. For the interpretation of the
observations, it is helpful to look at numerical simulations of wave
propagation through a magnetically structured solar model
atmosphere. In the past such simulations have always been carried out
for idealized atmospheric models and magnetic configurations
\citep{nutto_rosenthal02,nutto_bogdan03}. However, in reality the
magnetic field is expected to have a complex structure. 

Here we report on two-dimensional simulations of magneto-acoustic wave
propagation in a model of the solar atmosphere including realistic
granular velocity fields and complex magnetic fields.  In
Sect.~\ref{nutto_Sec:Method} we will give a short description of
the setup of the atmospheric model and of the excitation of the
waves. In Sect.~\ref{nutto_Sec:results} we will present and discuss
snapshots of our wave propagation simulations. Results are summarized
in the last section.

\section{Method}\label{nutto_Sec:Method}

For the simulations of magneto-acoustic wave propagation, we carry out
two-dimensional numerical experiments with the CO$^5$BOLD
code\footnote{see\,\textsf{www.astro.uu.se/\~{}bf/co5bold\_main.html}
  for the man pages of CO5BOLD}. The code solves the
magnetohydrodynamic equations for a fully compressible gas including
radiative transfer. Thus, the model atmosphere shows a realistic
granular velocity field and magnetic flux concentrations in the
intergranular lanes caused by the advective motion of the
granules. More details on the application of the the CO$^5$BOLD code
for wave propagation experiments can be found in
\citet{nutto_steiner07}.

\subsection{Two-dimensional model of the solar atmosphere}\label{nutto_Sec:two_dim}

The model atmosphere used for the two-dimensional numerical wave
propagation experiments is basically the same as was used by
\citet{nutto_steiner07}. The computational domain of the box covers
the upper layer of the convection zone from about $-1\,300$~km below
mean optical depth unity all the way to the middle layers of the
chromosphere at about 1\,600~km above mean optical depth unity. Along
the vertical direction an adaptive grid with 188 grid points is used
corresponding to cell sizes of 46~km for the biggest cells in the
convection zone and 7~km for the smallest cells in the
chromosphere. Transmitting boundary conditions are applied for the
lower and upper boundary. The lateral dimension of the box is
5\,000~km, with a grid resolution of 123 grid points corresponding to
a horizontal spatial resolution of 40~km. Periodic boundary
conditions are applied in the lateral directions.

\begin{figure}
\resizebox{\hsize}{!}{
\includegraphics[clip=true]{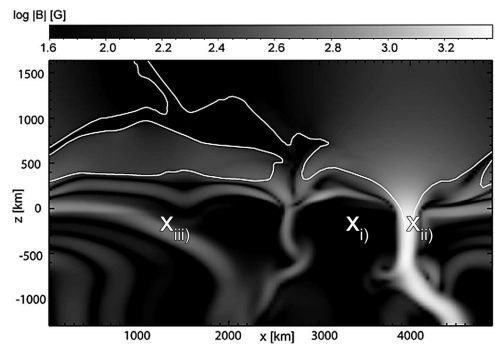}}
\caption{\footnotesize 
Magnetic setup of the model atmosphere represented in gray 
scales of $\log |B|$. The white curve corresponds to the 
equipartition layer. On the right hand side a flux sheet with 
strengths up to 2\,300~G has evolved. On the left hand side in the 
photosphere are weak horizontal magnetic fields. The crosses 
indicate the location of the wave excitation, where the numbers 
correspond to the three cases mentioned in the text.}
\label{nutto_Fig:2d_magsetup}
\end{figure}

Figure~\ref{nutto_Fig:2d_magsetup} gives an impression of the magnetic
setup.
The plot shows the absolute magnetic field strength $\log |B|$. Since
we are especially interested in the interaction of the waves with the
magnetic field, we use a model where a strong flux sheet has evolved
during the simulation time. In the core of the flux sheet, which is
present on the right hand side of the box, magnetic field strengths up
to 2\,300~G are reached. On the left-hand side weaker horizontal
fields have evolved. The white contour shows the equipartition layer
where the Alfv\'en speed equals the sound speed, and approximately
coincides with the $\beta =1$ layer. At this layer the acoustic and
magnetic mode of the magneto-acoustic waves can exchange energy
\citep{nutto_cally07}. This layer is known as the \emph{mode
  conversion zone}, where an acoustic wave can change to a magnetic
mode and vice versa. The strength of this interaction depends on the
angle between the wave vector and the magnetic field (attack
  angle), the wave number $k$, and the width of the conversion zone
\citep{nutto_cally07}. To investigate this interaction we excite waves
at three different positions in the computational domain. The
excitation positions are indicated by crosses in
Fig.~\ref{nutto_Fig:2d_magsetup}. The form of the excitation will be
discussed in Sect.~\ref{nutto_Sec:excitation}. The location of
the excitation is at a depth for which it was shown that the
excitation of waves in the Sun is most likely to take place
namely between the surface and 500~km in depth
\citep{nutto_stein01}. The three lateral positions are chosen at
locations where interesting cases of wave and magnetic field
interactions occur:
\begin{itemize}
\item[i)] where the magneto-acoustic waves interact with 
  the magnetic canopy of the flux sheet, 
\item[ii)] where the magneto-acoustic waves are excited inside the
  flux sheet, and
\item[iii)] where the magneto-acoustic waves interact with weak horizontal
  magnetic fields.
\end{itemize} 

\subsection{Wave excitation}\label{nutto_Sec:excitation}

In order to excite waves in our simulation domain we follow the
description given by \citet{nutto_parchevsky09}, where the spatial and
temporal behavior of the wave source is modeled by the function:
\begin{equation} 
  f(r,t)\!=\!\begin{cases}
    A
    \left[\!1\!-\!\frac{r^2}{R^2_{\mathrm{src}}}\!\right]^2\!\!(1-2\tau^2)\mathrm{e}^{-\tau^2}&\!\!\!\!\mbox{for }\, r \le R_{\mathrm{src}}, \\
    0&\!\!\!\!\mbox{for }\, r > R_{\mathrm{src}}, 
\end{cases}\label{nutto_Eq:rickerswavelet}
\end{equation}
where $$\tau=\frac{\omega_0(t-t_0)}{2}-\pi \quad \mbox{with} \quad
t_0\le t\le t_0 + \frac{4\pi}{\omega_0},$$ $R_{\mathrm{src}}$ is the
radius of the source, $A$ is the amplitude of the disturbance, and
$\omega_0=2\pi f_0$ determines the central frequency of the wave
excitation. This function can now be used as a source function in the
momentum equations by taking the gradient of
Eq.~\ref{nutto_Eq:rickerswavelet}, $\vec S= \nabla f$. Then, the
components of the source function $\vec S$ can be combined with the
components of the pressure gradient. Thus, the source function can be
seen as a perturbation to the pressure gradient, launching an acoustic
wave with a central frequency of $\omega_0$. 

\begin{figure}
\resizebox{\hsize}{!}{
\includegraphics[clip=true]{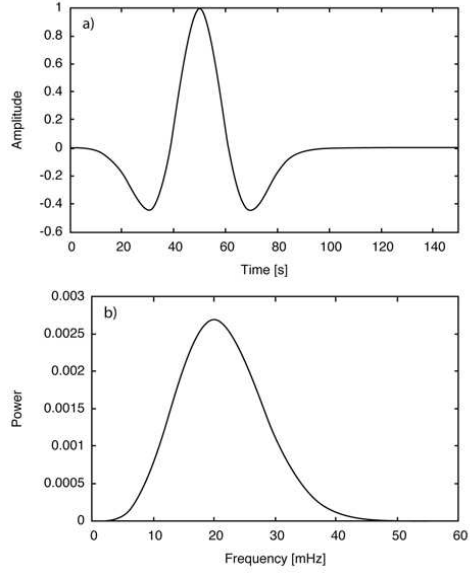}}
\caption{\footnotesize 
a) Temporal behavior of the wave source. b) Power of the wave 
source. The pressure disturbance excites a non-monochromatic wave 
with a central frequency of $f_0=20$~mHz.}
\label{nutto_Fig:excitation} 
\end{figure}

For our simulations we use a central frequency of $f_0=20$~mHz and a radius
of the source of $R_\mathrm{src}=50$~km. The temporal behavior of the
wave source is plotted in panel a) of
Fig.~\ref{nutto_Fig:excitation}. The panel b) shows the power spectrum
of the excitation source. A non-monochromatic wave is excited with
most of the power around $f_0=20$~mHz.

\section{Results}\label{nutto_Sec:results}

In this section we present snapshots taken from our simulations of
wave propagation in a complex magnetically structured atmosphere. We
will focus on the interaction between the wave and the magnetic field
that is taking place in the mode conversion zone.

\begin{figure*}
\center
\resizebox{0.9\hsize}{!}{
\includegraphics[clip=true]{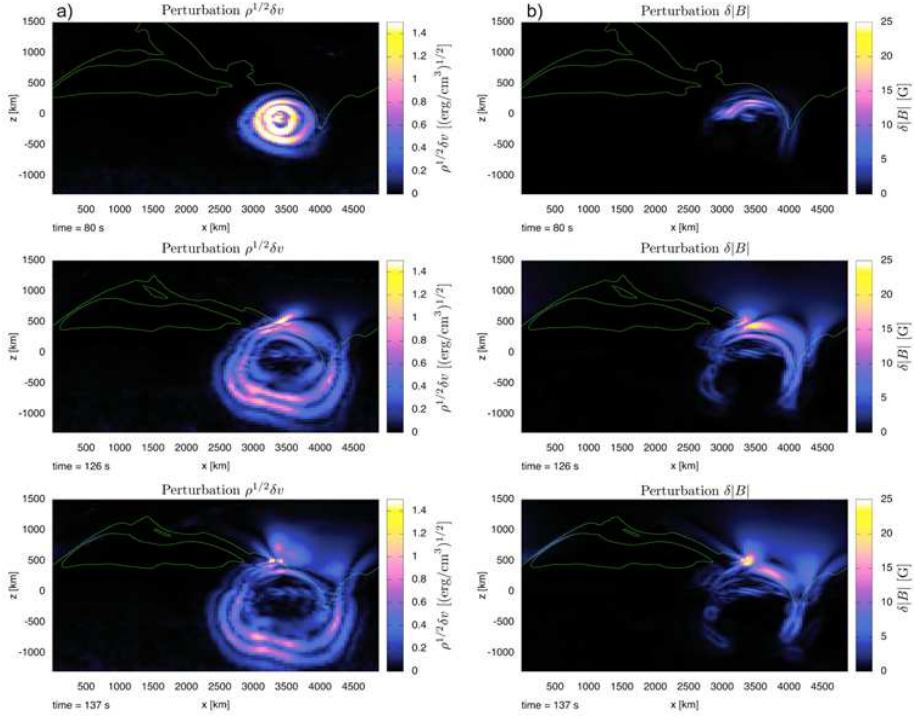}}
\caption{\footnotesize 
a) Perturbation of the square root of the energy density, 
$\rho^{1/2}\delta v$, caused by a wave that is excited outside in the 
vicinity of the flux sheet. Shown are three snapshots taken at 
80~s, 126~s, and 137~s into the simulation run. b) 
Perturbation $\delta B$ of the magnetic field. The snapshots are 
taken at the same time as in panel a). In all plots, the solid 
contour corresponds to the equipartition layer indicating the 
conversion zone.}
\label{nutto_Fig:fluxsheet} 
\end{figure*}

Panel a) of Fig.~\ref{nutto_Fig:fluxsheet} shows snapshots 
of the propagation of a
wave that is excited just outside of a flux sheet underneath the
magnetic canopy. The propagation of both wave modes, the predominantly
acoustic and predominantly magnetic mode, can be tracked in the
perturbation of the square root of the energy density
$\delta\epsilon^{1/2}=\rho^{1/2}\delta v$. Because of the position of the
wave source beneath the $\beta=1$ layer, in a region of high plasma
$\beta$, the applied wave source excites a wave that is predominantly
an acoustic mode. Upon reaching the conversion zone, indicated by the
solid contour in Fig.~\ref{nutto_Fig:fluxsheet}, energy is
transferred from the acoustic to the magnetic mode
\citep{nutto_cally07}. Because of the large attack angle between the
local wave vector and the magnetic field vector, most of the energy is
converted to the fast magnetic mode. Transmission of the acoustic mode
is not apparent. The magnetic mode is best seen in the perturbation
$\delta B$ of the magnetic field, plotted in panel b) of
Fig.~\ref{nutto_Fig:fluxsheet}. The second plot in panel b), at
126~s into the simulation, shows the fast magnetic mode emerging
from the conversion zone. Because of the gradient of the Alfv\'en speed,
the wave is refracted back down into the photosphere.

\begin{figure*}
\center
\resizebox{0.9\hsize}{!}{
\includegraphics[clip=true]{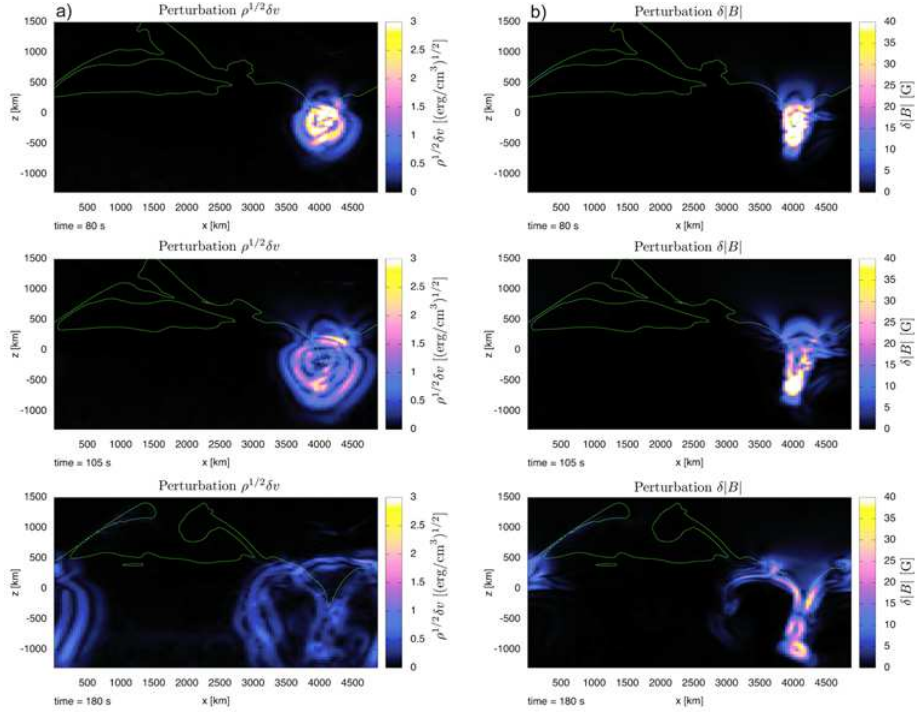}} 
\caption{\footnotesize 
a) Perturbation $\delta\epsilon^{1/2}$ caused by a wave that is 
excited inside of the flux sheet close to the equipartition 
layer. Shown are three snapshots taken at 80~s, 105~s, and 
180~s into the simulation run. b) Perturbation $\delta B$ of the 
magnetic field. The snapshots are taken at the same time as in panel 
a). In all plots, the solid contour corresponds to the 
equipartition layer indicating the wave-conversion zone.}
\label{nutto_Fig:inB} 
\end{figure*}

Next, we will focus on the wave propagation when the wave source is
located inside the flux sheet close to the equipartition layer. Figure
\ref{nutto_Fig:inB} shows again snapshots of the perturbation of the
square root of the energy density $\delta\epsilon^{1/2}$, panel a),
and the perturbation $\delta B$ of the magnetic field, panel b). Since
the wave source is located above the equipartition layer in the
low-$\beta$ regime, the fast magnetic wave mode is excited
directly and together with the slow acoustic mode, different from the
previous case.
Because of the higher Alfv\'en speed above the equipartition layer, the
fast magnetic wave runs ahead of the slow acoustic mode. This fast
magnetic mode can again be easily identified in the perturbation
$\delta B$ of the magnetic field, plotted in panel b) of
Fig.~\ref{nutto_Fig:inB}. As in the previous case, the gradient of the
Alfv\'en speed causes the fast magnetic wave to refract to a degree
that it travels back towards the lower atmosphere. The slow acoustic
wave emerges from the location of the wave excitation at a later time
(last plot in panel a) of Fig.~\ref{nutto_Fig:inB}), and is then guided
along the magnetic field lines into the higher layer of the
atmosphere. As the acoustic wave travels up the atmosphere the wave
front is steepening and starts to shock.

\begin{figure*}
\center
\resizebox{0.9\hsize}{!}{
\includegraphics[clip=true]{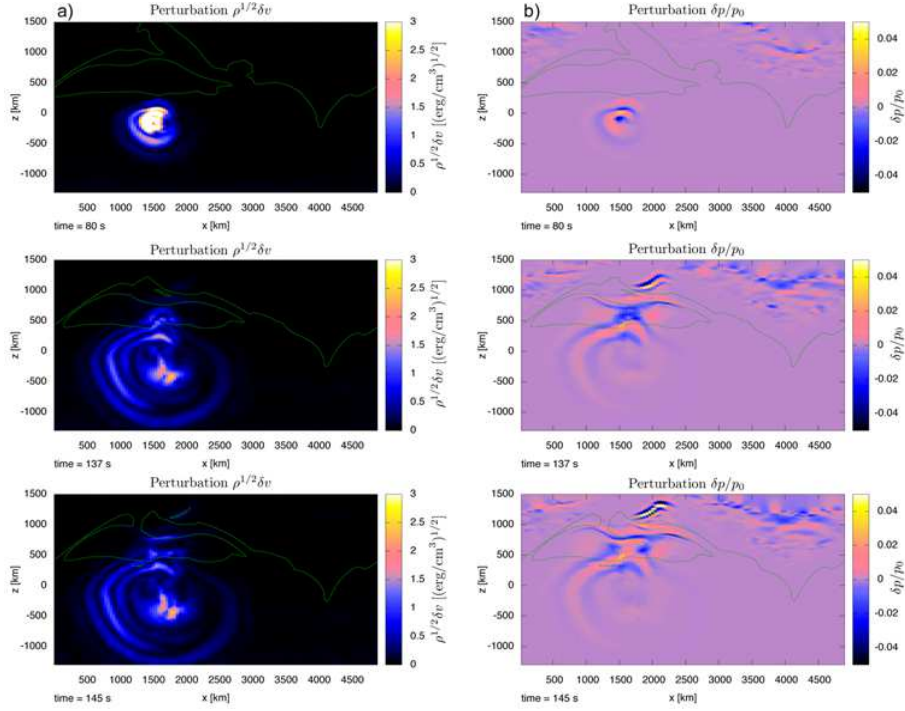}}
\caption{\footnotesize 
a) Perturbation $\delta\epsilon^{1/2}$ caused by a wave that is 
excited beneath weak horizontal magnetic fields. Shown are three 
snapshots taken at 80~s, 137~s, and 145~s into the 
simulation run. b) Perturbation of the pressure, $\delta p/p_0$. The 
snapshots are taken at the same time as in panel a). In all plots, 
the solid contour corresponds to the equipartition layer 
indicating the wave-conversion zone.}
\label{nutto_Fig:voidB} 
\end{figure*}

Finally, we excite a wave just below weak horizontal magnetic
fields. Snapshots of the wave propagation are plotted in
Fig.~\ref{nutto_Fig:voidB}, where panel a) shows again the
perturbation of the square root of the energy density
$\delta\epsilon^{1/2}$. Panel b) shows the perturbation of the
pressure $\delta p/p_0$, with $p_0$ being the local unperturbed
pressure.
As the excitation location is below the equipartition layer in the
high-$\beta$ regime, primarily the fast acoustic wave mode is
excited. At the first equipartition layer, the local wave and magnetic
vector are almost perpendicular to each other. Thus, most of the
acoustic wave energy is converted into the fast magnetic mode. A
transmitted acoustic wave is not visible at first. However, the
existence of multiple equipartition layers in the path of the
magneto-acoustic waves, adds to the complexity of the
interpretation. Multiple equipartition layers have never been
addressed in former investigations. Due to the large attack angle, one
expects that no acoustic wave should be transmitted through the mode
conversion zone. However, by looking at the pressure disturbance
$\delta p/p_0$ in Fig.~\ref{nutto_Fig:voidB} it can be seen that an
acoustic wave can be transmitted through the multiple equipartition
layers by a series of mode conversion. In the middle plot of panel b)
a strong pressure disturbance starts to reemerge above the multiple
conversion zones. This can be seen as an indication of a transmitted
acoustic wave in spite of the horizontal magnetic fields.

\section{Summary}

We have shown numerical experiments of wave propagation in a realistic
two-dimensional model atmosphere. We focused on the interaction
between acoustic and magnetic wave modes at the equipartition layer
where the Alfv\'en speed is similar the sound speed. 

When the location of the wave source is below the equipartition layer
in the high-$\beta$ regime, the launched wave is predominantly
acoustic in nature. In the first and third of the presented cases, the
large attack angle of the local wave vector with the magnetic
field vector results in the conversion of the fast acoustic mode into
the fast magnetic mode. When there are multiple equipartition layers,
an acoustic mode can be transmitted into the higher layers of the
atmosphere by a series of mode conversions. In all three cases
the fast magnetic mode is usually refracted back towards the lower
atmosphere because of the gradient in Alfv\'en speed.

If the location of the wave
source is close to the equipartition layer, both, the acoustic and the
magnetic mode, are excited. The slow acoustic wave trails the fast
magnetic wave. While the latter shows an upper turning point again,
the slow acoustic wave is guided along the magnetic field into the
higher layers of the atmosphere where the wave starts to
shock.

The simulations confirm that the propagation of
magneto-acoustic waves in the solar atmosphere is strongly influenced
by the equipartition layer. In comparison to investigations where
static background models are used, our simulations demonstrate that
the equipartition layer can show a complex topography and it is highly
dynamic itself.

\begin{acknowledgements}
The authors acknowledge support from the European Helio- and
Asteroseismology Network (HELAS), which is funded as Coordination Action
by the European Commission's Sixth Framework Programme. C. Nutto thanks
NSO for the support of a travel grant to visit the NSO Workshop \# 25.
\end{acknowledgements}

\bibliographystyle{aa}

\end{document}